\documentclass[12pt]{article}
\usepackage[english]{babel}
\usepackage{geometry,amsmath,amssymb,graphicx,hyperref}
\newcommand{\mathe}{\mathrm{e}}
\newcommand{\nobracket}{}

\newcommand{\tmem}[1]{{\em #1\/}}
\newcommand{\tmop}[1]{\ensuremath{\operatorname{#1}}}
\newcommand{\tmtextbf}[1]{{\bfseries{#1}}}
\newcommand{\tmtextit}[1]{{\itshape{#1}}}
\newcommand{\tmtexttt}[1]{{\ttfamily{#1}}}
\newenvironment{itemizedot}{\begin{itemize} }{\end{itemize}}
\begin{document}

\title{Ising model as Wilson-Majorana Fermions}

\author{Ulli Wolff\thanks{
e-mail: uwolff@physik.hu-berlin.de} \\
Institut f\"ur Physik, Humboldt Universit\"at\\ 
Newtonstr. 15 \\ 
12489 Berlin, Germany
}
\date{}
\maketitle

\begin{abstract}
  We show the equivalence of the 2D Ising model to standard free Euclidean
  lattice fermions of the Wilson Majorana type. The equality of the loop
  representations for the partition functions of both systems is established
  exactly for finite lattices with well-defined boundary conditions. The
  honey\-comb lattice is particularly simple in this context and therefore
  discussed first and only then followed by the more familiar square lattice
  case.
\end{abstract}

\

\section{Introduction}

The two-dimensional Ising spin model can probably be called {\tmem{the}}
prototype exactly solved model in statistical physics and lattice field
theory. The break-through solution was achieved by Onsager
{\cite{PhysRev.65.117}} who has computed the partition function by the
transfer matrix approach in 1944. Since then a large number of alternative --
usually less complicated -- strategies to derive the same result have been
presented, like for instance {\cite{Kaufman:1949ks}}. Without any attempt
toward completeness\footnote{A more extended recent collection of references
is found in {\cite{Strecka}} for example. See also {\cite{mccoy1973two}}.} we
shall only mention below a few more selected references which are more or less
close to our approach. Finally, we then hopefully will be able to sufficiently
justify the present addition to this literature.

On a very naive level one may find the two-valuedness of the elementary spins
reminiscent of fermionic states that can be empty and occupied. A much more
concrete such link was established in the paper of Schultz, Mattis and Lieb
{\cite{Schultz:1964fv}}. The transfer matrix of the model is first expressed
in terms of a tensor product of Pauli algebras attached to the sites of a one
dimensional row of the lattice. By performing a Jordan-Wigner transformation
{\cite{Jordan:1928wi}} the Pauli matrices are traded for anticommuting fermion
operators. The transfer matrix in this form inherits the nearest neighbor
bilinear structure of the original model and can hence be diagonalized by
Fourier expansion on the lattice. Simpler analogous steps in the Hamiltonian
limit of the transfer matrix (`continuous time') have later been discussed in
{\cite{PFEUTY197079}} and {\cite{Kogut:1979wt}}.

An alternative to representing fermions by operators with canonical
anticommutation relations is given by path integrals over anticommuting
Grassmann `numbers' {\cite{Berezin}}. Such a representation has been derived
from the operator transfer matrix in {\cite{Itzykson:1982ed}}. Even earlier,
Samuel {\cite{Samuel:1978zx}} has used Grassmann integrals to directly `draw'
the high temperature series of the Ising model to all orders which constitutes
an equivalent representation of the model. In either case the resulting
integrals are Gaussian (free fermions), can be performed, and thus furnish an
exact solution.

Samuel's work is by far the closest to our work. The differences
distinguishing the presentation at hand are however the following. We
demonstrate the equivalence of the Ising model with free Euclidean Majorana
fermions of the Wilson type {\cite{Wilson:1975id}} that is one of the standard
choices in lattice field theory. The critical point corresponds to zero mass
and the Euclidean relativistic invariance in the continuum/scaling limit is
manifest in this standard framework. All phase factors in the matched
expansions arise naturally from the fermion nature combined with half-angle
spin rotation phases. The Fermi-Bose equivalence proven here will be an exact
identity between arbitrary finite lattice partition functions with
well-defined periodic or antiperiodic boundary conditions in each of the two
lattice directions.

In section \ref{sec2} the loop representation of the Ising model is defined
and matched to the fermionic model for the honeycomb lattice. In section
\ref{sec3} the same program is implemented for the standard square lattice
which is technically more complicated. Some conclusions and remarks on
more than two dimensions are offered in section \ref{sec4}. In two appendices
we report details on the evaluation of the spin weights and on the actual
evaluation of the free fermion partition functions. In particular, the lattice
fermion spectra are plotted.

\section{Equivalence on the honeycomb lattice}\label{sec2}

\subsection{Honeycomb geometry}

\

\begin{figure}[htb]
\begin{center}
  \resizebox{0.9\textwidth}{!}{\includegraphics{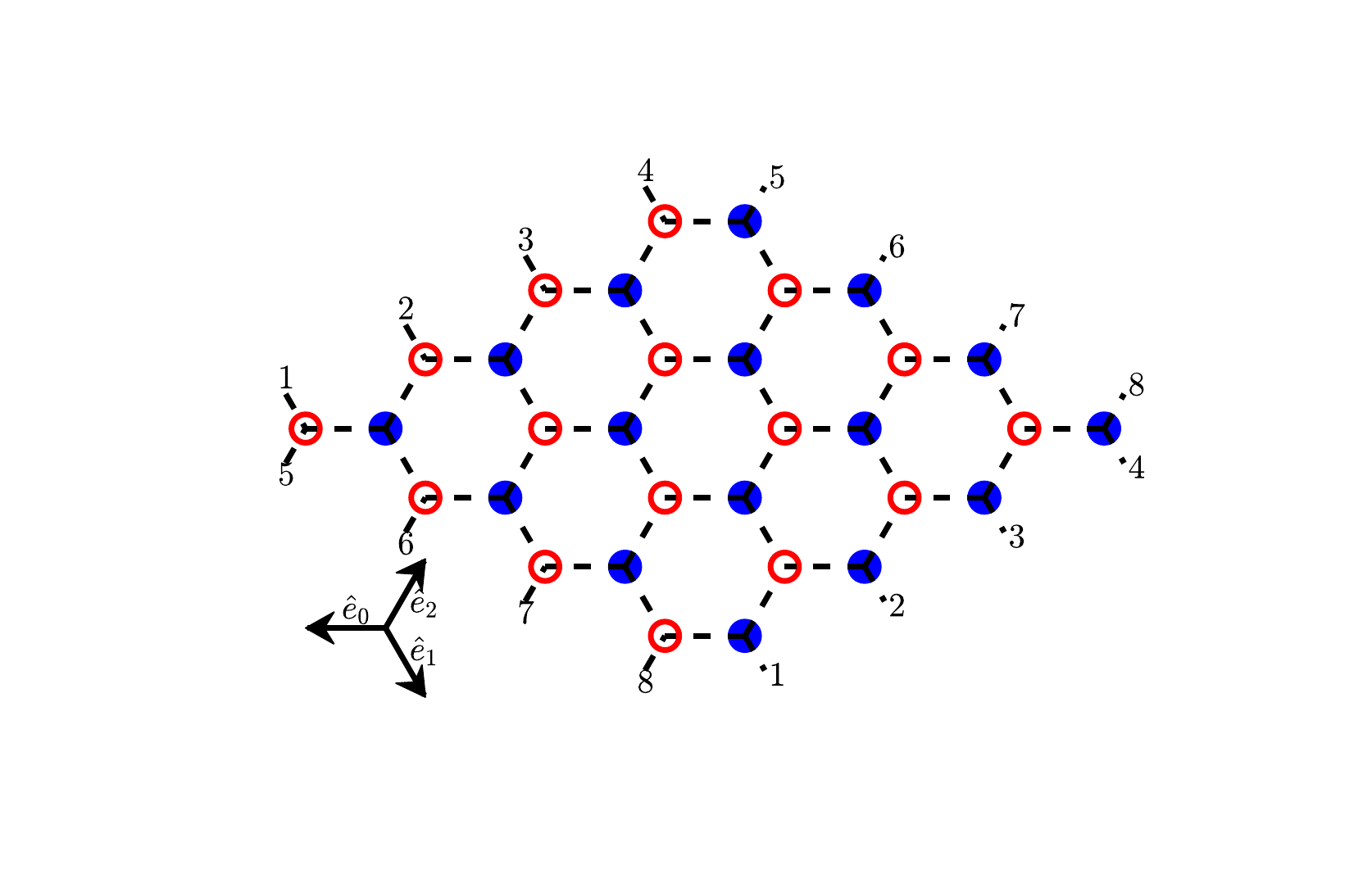}}
  \caption{The honeycomb lattice with $L_1 = L_2 = 4$. Sites of
  $\mathcal{A}$($\mathcal{B}$) carry full blue (empty red) dots. Periodic
  boundary conditions are indicated by open links with equal
  indices.\label{hgeo}}
\end{center}
\end{figure}

The honeycomb lattice can be spanned by two triangular sublattices
$\mathcal{A}$ and $\mathcal{B}$, see figure \ref{hgeo}. Each site $x \in
\mathcal{A}$ has three nearest neighbors $x + \hat{e}_a \in \mathcal{B}$, $a =
0, 1, 2$ where the three unit vectors\footnote{We use lattice units $a = 1$
with respect to these nearest neighbor links.} making 120 degree angles with
each other fulfill
\begin{equation}
  \hat{e}_a \cdot \hat{e}_b = \frac{1}{2}  (3 \delta_{a b} - 1) . \label{esp}
\end{equation}
The sites of $\mathcal{A}$ are labeled by integers $x_1, x_2$ in the form
\begin{equation}
  \mathcal{A} \ni x = x_1 f_1 + x_2 f_2, \quad f_1 = \hat{e}_1 - \hat{e}_0,
  \quad f_2 = \hat{e}_2 - \hat{e}_0, \quad f_i \cdot f_j = \frac{3}{2} 
  (\delta_{i j} + 1) . \label{fsp}
\end{equation}
All sites in $\mathcal{B}$ can now be generated as neighbors $x + \hat{e}_0$
of a unique $x \in \mathcal{A}$.

A simple way to impose periodic boundary conditions\footnote{In
{\cite{Smith:2014tha}} a more `rectangular' periodicity is introduced that
results in helical boundary conditions for the honeycomb fields. Clearly, also
this case can be handled along the lines presented here.} to obtain a finite
system is to identify points $x$ with $x + L_1 f_1$ and $x + L_2 f_2$ with
integer $L_i$. We then have $V = 2 L_1 L_2$ independent sites in total, half
in $\mathcal{A}$ and half in $\mathcal{B}$.

We adopt the convention to take coordinates $x_i \in [0, L_i)$ and label links
by the pairs $(x, a)$ with $x \in \mathcal{A}$, $a \in \{0,1,2\}$. Which links close `around the
boundary'? The neighbors $x + \hat{e}_a$ are in $\mathcal{B}$ and in our
coordinate system associated with (generated by) $x + \hat{e}_a - \hat{e}_0$
back in $\mathcal{A}$, or in other words, with $x, x + f_1, x + f_2$ which are
folded back into the ranges $[0, L_i)$ by standard modulo operations. A site
$y \in \mathcal{B}$ on the other hand is generated by $y - \hat{e}_0 = z \in
\mathcal{A}$ and its neighbors $y - \hat{e}_a$ are $z, z - f_1, z -
f_2$.\tmtextit{}

\

\subsection{Ising model}\label{IsingMajo}

We attach Ising spins $s (x) \in \{ + 1, - 1 \}$ to all sites and write the
partition function of the Ising model on the honeycomb lattice as
\begin{equation}
  Z = \sum_s \mathe^{\beta \sum_{a, x \in \mathcal{A}} \, s (x) s (x +
  \hat{e}_a)} = 2^V (\cosh \beta)^{3 V / 2} Z_r .
\end{equation}
The reduced partition function $Z_r$ is given by
\begin{equation}
  Z_r = 2^{- V} \sum_s \prod_{a, x \in \mathcal{A}} (1 + t s (x) s (x +
  \hat{e}_a)), \qquad t \equiv \tanh \beta . \label{Zrfact}
\end{equation}
We read off the loop graph representation of $Z_r :$
\begin{itemizedot}
  \item we multiply out the big product and for each term draw lines on the
  links where the $t s s$ term is picked and leave empty links with factors
  one,
  
  \item after averaging each term over $s$, nonzero contributions arise only
  from graphs where each site is surrounded by an even number of lines,
  
  \item as each site has only 3 neighbors only zero or two lines are allowed
  at sites,
  
  \item therefore each nonzero contribution to $Z_r$ can be seen as
  configuration of multiple non-intersecting closed loops,
  
  \item $Z_r$ is given by the sum over all different such loop gas
  configurations weighted with a factor $t$ per line segment.
\end{itemizedot}
Symbolically we may write
\begin{equation}
  Z_r = \sum_{\Lambda} t^{| \Lambda | \nobracket \nobracket}, \label{Zrsum}
\end{equation}
where the sum runs over the loop gas configurations on the lattice and $|
\Lambda | \nobracket \nobracket$ means the total number of links making up all
the closed loops contained in $\Lambda$.

Up to here we have tacitly assumed periodic boundary conditions. For either or
both of the directions $f_1, f_2$ in which we close the torus, we may also
take antiperiodic boundary conditions. We designate the 4 possible cases that
arise by bits $\varepsilon_1, \varepsilon_2$ with $\varepsilon_i = 0$ standing
for periodic and $\varepsilon_i = 1$ for antiperiodic. To detect if
antiperiodicity leads to negative amplitudes we define winding numbers
\begin{equation}
  q_i [\Lambda] = \tmop{number} \tmop{of} \tmop{occupied} \tmop{links} (x, i)
  \left|_{x_i = L_i - 1}  (\tmop{mod} 2) . \label{qidef} \right.
\end{equation}
Then, for generalized boundary conditions, the sign $(- 1)^{\varepsilon_1 q_1
+ \varepsilon_2 q_2}$ has to be included in the sum in (\ref{Zrsum}). The
generalized Ising partition function with dynamical boundary conditions, in
which we sum over the four cases with weights $\rho (\varepsilon)$, reads
\begin{equation}
  Z_{I \rho} = \sum_{\Lambda} t^{| \Lambda | \nobracket \nobracket}
  \Phi_{\rho} [\Lambda] \quad \tmop{with} \quad \Phi_{\rho} [\Lambda] =
  \sum_{\varepsilon} \rho (\varepsilon) (- 1)^{\varepsilon_1 q_1 [\Lambda] +
  \varepsilon_2 q_2 [\Lambda]} . \label{Zrho}
\end{equation}
This enlarged ensemble, including a sum over boundary conditions, will be
found to be a convenient starting point for the finite size equivalence to be
derived.

\subsection{Majorana Wilson fermion}

We consider a two component Grassmann-valued field $\xi_{\alpha} (x), \alpha =
1, 2$ living on the sites of the honeycomb lattice. It is endowed with the
Gaussian Euclidean action
\begin{equation}
  S = \frac{1}{2} \sum_x \bar{\xi} (x) \xi (x) - \kappa \sum_{a, x \in
  \mathcal{A}} \bar{\xi} (x) P (\hat{e}_a) \xi (x + \hat{e}_a) \label{Smajo}
\end{equation}
written in the hopping parameter form. For a unit vector $n$, $P (n)$ is the
Wilson projector
\begin{equation}
  P (n) = \frac{1}{2} (1 - n_{\mu} \gamma_{\mu}) . \label{Wpro}
\end{equation}
The $2 \times 2$ Dirac matrices generate the Clifford algebra
\begin{equation}
  \{ \gamma_{\mu}, \gamma_{\nu} \} = 2 \delta_{\mu \nu} .
\end{equation}
Note that $\mu, \nu = 0, 1$ refer to a pair of orthogonal directions in the
plane. Due to the Majorana nature of $\xi$ the field $\bar{\xi} (x)$ is not
independent but given by
\begin{equation}
  \bar{\xi} = \xi^{\top} \mathcal{C}
\end{equation}
with the charge conjugation matrix $\mathcal{C}$ defined by
\begin{equation}
  \gamma_{\mu}^{\top} = -\mathcal{C} \gamma_{\mu} \mathcal{C}^{- 1} .
  \label{Ctrafo}
\end{equation}
In any representation one may prove antisymmetry, $\mathcal{C}= -
\mathcal{C}^{\top}$, and our normalization will be $\mathcal{C}_{12} = + 1$.
We note that (\ref{Smajo}) contains a simple sum over all links. They are
unoriented -- as in the Ising model -- because
\begin{equation}
  \bar{\xi} (x) P (\hat{e}_a) \xi (x + \hat{e}_a) = \bar{\xi} (x + \hat{e}_a)
  P (- \hat{e}_a) \xi (x)
\end{equation}
holds due to (\ref{Ctrafo}).

To study the so called naive continuum limit, we substitute\\
$\xi (x + \hat{e}_a) \simeq (1 + \hat{e}_a \cdot \partial) \xi (x)$. Using the
identities
\begin{equation}
  \sum_a \hat{e}_a = 0, \quad \sum_a \hat{e}_{a, \mu}  \hat{e}_{a, \nu} =
  \frac{3}{2} \delta_{\mu \nu}
\end{equation}
we find
\begin{equation}
  S \simeq \frac{3 \kappa}{4}  \sum_{x \in \mathcal{A}} \bar{\xi}
  (\gamma_{\mu} \partial_{\mu} + m) \xi \quad \tmop{with} \quad m =
  \frac{2}{\kappa} (2 / 3 - \kappa) . \label{mkappa}
\end{equation}
Hence, by rescaling $\xi$ we have a canonical Majorana fermion of mass $m$. A
small positive mass (in lattice units) appears as the hopping parameter
$\kappa$ approaches the critical value $\kappa_c = 2 / 3$ from below. A
discussion of the complete dispersion relation in momentum space and the exact
partition function is given in appendix \ref{appb1}

The fermion partition function is given by the Grassmann integral
\begin{equation}
  Z_M = \int D \xi \mathe^{- S} = \int D \xi \left\{ \prod_x \left( 1 -
  \frac{1}{2} \bar{\xi} \xi \right) \right\} \prod_{a, x \in \mathcal{A}} [1 +
  \kappa \bar{\xi} (x) P (\hat{e}_a) \xi (x + \hat{e}_a)] . \label{Zfact}
\end{equation}
The integration over two Grassmann components per site factorizes $D \xi =
\prod_x d^2 \xi$ and the local measure $d^2 \xi$ is taken such that
\begin{equation}
  \int d^2 \xi \, \xi_{\alpha} \bar{\xi}_{\beta} = \delta_{\alpha \beta} \quad
  \Rightarrow \quad \int d^2 \xi (-) \frac{1}{2} \bar{\xi} \xi = 1.
  \label{Elgrass}
\end{equation}
Moreover, to arrive at the factorized form (\ref{Zfact}), the nilpotency of
the Grassmann bilinears has been used, including the fact that $P (\hat{e}_a)$
are one-dimensional projectors.

A moment of thought will reveal now, that upon executing the Grassmann
integrations site by site and using (\ref{Elgrass}) the same loop gas
structure arises as from (\ref{Zrfact}). For each loop $\lambda$ the successive
projectors $P$ appear multiplied up and an over-all factor
\begin{equation}
  w (\lambda) = - \tmop{tr} [P (n_1) P (n_2) \cdots P (n_N)] . \label{wdef}
\end{equation}
arises with $n_1, n_2, \ldots, n_N$ being the unit vectors $\pm \hat{e}_a$ met
on the links along the loop. The minus sign is the usual fermionic one: We
order the commuting bilinears for the sequence of links in a loop
schematically as $\bar{\xi} P \xi \, \bar{\xi} P \xi \cdots
\bar{\xi} P \xi$. Successive inner pairs $\xi \, \bar{\xi}$ are at
the same site and integrate to $\delta_{\alpha \beta}$. The first
$\bar{\xi}$ and the last $\xi$ similarly close the trace but come in the
`wrong' order, hence a factor $- 1$. The geometric factor $w$ is evaluated in
detail in appendix \ref{appa}. For periodic boundary conditions in both
directions, for example, we are led to
\begin{equation}
  Z_M = \sum_{\Lambda} [\kappa \cos (\pi / 6)]^{| \Lambda | \nobracket
  \nobracket} (2 \delta_{q_1 [\Lambda], 0} \delta_{q_2 [\Lambda], 0} - 1) .
\end{equation}
Some further explanations are in order:
\begin{itemize}
  \item Closed loops on the honeycomb lattice have as many 60 degree bends as
  they contain links. This results in the same powers of $\kappa$ per link and
  $\cos (\pi / 6)$ per bend (half-angle between $n_i$ and $n_{i + 1}$, see
  (\ref{weval})).
  
  \item For loops not winding around the torus, each Fermi sign is
  paired\footnote{It is this pairing that in our language makes two
  dimensional fermions special with an essentially positive loop
  representation.} with the spin minus from $2 \pi$ rotation (see appendix
  \ref{appa}). If $\Lambda$ contains loops winding around one or both periodic
  directions (nonzero $q_i$), the rotation is lacking and a minus sign is
  left.
\end{itemize}
For the superposition of boundary conditions with weight $\eta (\varepsilon)$
the Majorana partition function becomes
\begin{equation}
  Z_{M \eta} = \sum_{\Lambda} [\kappa \cos (\pi / 6)]^{| \Lambda | \nobracket
  \nobracket} \Phi_{M \eta} [\Lambda] \label{Zfrho}
\end{equation}
with
\begin{equation}
  \Phi_{M \eta} [\Lambda] = (2 \delta_{q_1 [\Lambda], 0} \delta_{q_2
  [\Lambda], 0} - 1) \sum_{\varepsilon} \eta (\varepsilon) (-
  1)^{\varepsilon_1 q_1 [\Lambda] + \varepsilon_2 q_2 [\Lambda]} .
\end{equation}
We see that this loop gas coincides with (\ref{Zrho}) if the following
matching conditions hold
\begin{equation}
  \tanh \beta = t = \kappa \cos (\pi / 6) = \frac{\sqrt{3}}{2} \kappa
  \label{matchbeta}
\end{equation}
and
\begin{equation}
  \Phi_{\rho} [\Lambda] = \Phi_{M \eta} [\Lambda] . \label{matchrho}
\end{equation}
The $\Phi_{\times}$ depend on the graph $\Lambda$ only through the winding
numbers $q_i [\Lambda]$ and their equality translates into a relation between
$\rho$ and $\eta$ as follows. We may view $\varepsilon_i$ and $q_i$ as
conjugate binary Fourier variables and invert
\begin{equation}
  \rho (\varepsilon) = \frac{1}{4}  \sum_q \Phi_{\rho} (q) (-
  1)^{\varepsilon_1 q_1 + \varepsilon_2 q_2} .
\end{equation}
If we impose (\ref{matchrho}) this implies
\begin{equation}
  \rho (\varepsilon) = \frac{1}{4}  \sum_q (2 \delta_{q_1, 0} \delta_{q_2, 0}
  - 1) \sum_{\varepsilon'} \eta (\varepsilon') (- 1)^{\varepsilon'_1 q_1 +
  \varepsilon'_2 q_2} = 2 \bar{\eta} - \eta (\varepsilon), \hspace{1.2em}
  \bar{\eta} = \frac{1}{4} \sum_{\varepsilon} \eta (\varepsilon)
\end{equation}
or the particularly symmetric form
\begin{equation}
  \rho (\varepsilon) + \eta (\varepsilon) = 2 \bar{\rho} = 2
  \bar{\eta} .
\end{equation}
By setting for example $\rho (\varepsilon) = \delta_{\varepsilon,
\varepsilon'}$ we obtain\footnote{Here the normalization of $Z$ matters; we
fix it by demanding $Z_{\times} = 1$ for $\beta = \kappa = 0.$} for fixed
boundary conditions (for either the Ising or the Majorana system)
\begin{equation}
  Z_I (\beta, L_i, \varepsilon) + Z_M (\kappa, L_i, \varepsilon) = \frac{1}{2}
  \sum_{\varepsilon} Z_I (\beta, L_i, \varepsilon) = \frac{1}{2}
  \sum_{\varepsilon} Z_M (\kappa, L_i, \varepsilon) \label{Zrel}
\end{equation}
with $\beta, \kappa$ related by (\ref{matchbeta}). Obviously, by taking
derivatives, we may relate internal energy, susceptibility, etc. We have
checked our formulas by exact summation on some small lattices. We remark that
all $Z_{\times}$ here are even in $\beta$ or $\kappa$. This is shown by
flipping the signs for all fields on one of the two sublattices.

Combining (\ref{matchbeta}) with (\ref{mkappa}) the fermion mass (close to
crititicality) reads
\begin{equation}
  m = 2 \frac{t_c - t}{t}, \qquad t_c = \frac{1}{\sqrt{3}}, \qquad \beta_c =
  \frac{1}{2} \ln \left( 2 + \sqrt{3} \right) .
\end{equation}
Needless to say, the critical coupling of the Ising model on a honeycomb
lattice has been well-known before, see references in {\cite{Strecka}}. We see
that here the phase with $\kappa > \kappa_c$ or $m < 0$ of the free Wilson
fermion corresponds to the magnetized Z(2) symmetry-broken Ising phase.

\section{Equivalence on the square lattice}\label{sec3}

The Ising model is clearly most popular on the square lattice that we discuss
now. It will turn out, however, that the loop representation and the Majorana
form is a bit more complicated.

\subsection{Ising model}\label{Isquare}

The formulas analogous to those in section \ref{IsingMajo} are rather obvious
so that we here start immediately from the loop gas form which looks identical
to (\ref{Zrsum}) and (\ref{Zrho}) with just a re-definition of $\Lambda$. As
before $\Lambda$ is an arbitrary collection of line-carrying links such that
an {\tmem{even}} number of lines touch any site of the torus. This allows for
zero, two and, in contrast to the honeycomb lattice, also four links around a
site. Because of the latter possibility, to be called crossings from here on,
the configuration does in general not decompose into simple disjoint loops. By
some abuse of language it is however customary to still talk about a loop gas
configuration. The definition (\ref{qidef}) can also be taken over unchanged
if we substitute the orthogonal directions $\mu = 0, 1$ for $i = 1, 2$ and
$q_{\mu} [\Lambda]$ now are the corresponding modulo two winding numbers.

\subsection{Majorana Wilson fermion}

\begin{figure}[htb]
\begin{center}
  \resizebox{0.5\textwidth}{!}{\includegraphics{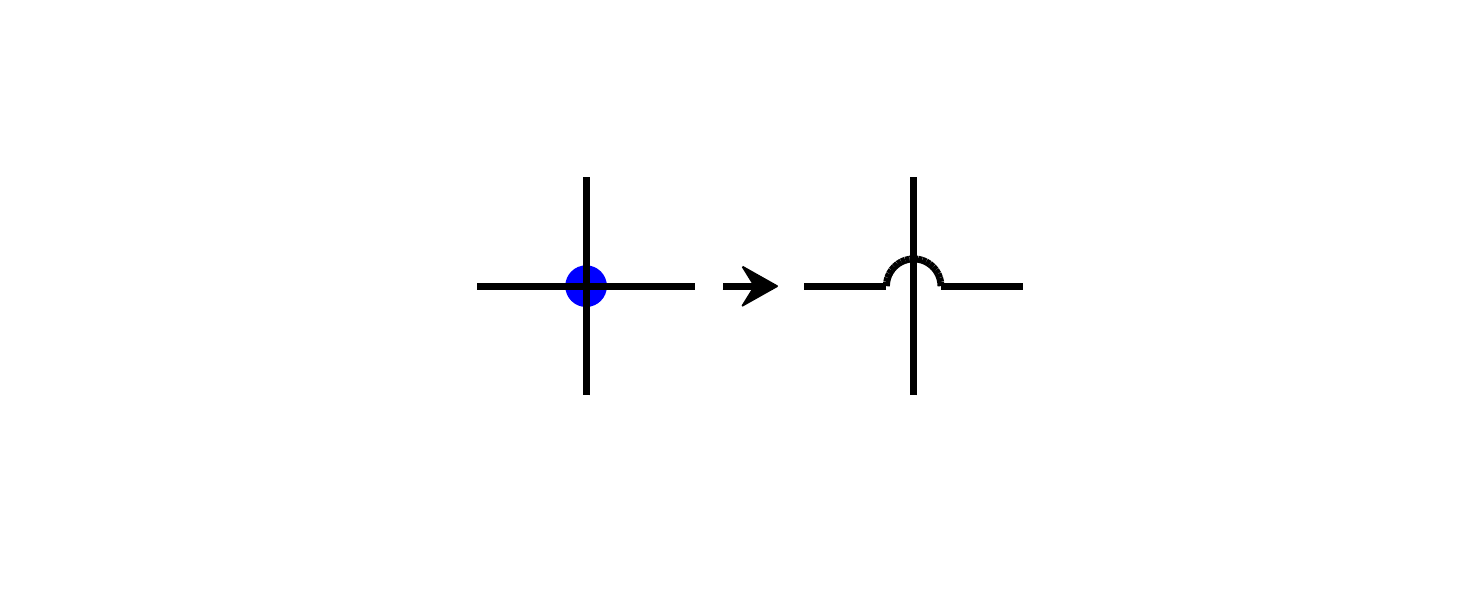}}
  \caption{Visualization of the spin contractions at crossings.\label{cross}}
\end{center}  
\end{figure}

The loop gas of a single species of Majorana Wilson fermions on the square
lattice has been discussed in {\cite{Wolff:2008xa}}. Attempting to `draw' the
Ising loop gas we notice two problems:
\begin{itemize}
  \item crossings cannot occur with only two Grassmann components per site,
  
  \item 90 degree bends come with weight factors $\cos (\pi / 4) = 1 /
  \sqrt{2}$ and their total power \ is not simply determined by the number of
  links as for the previous lattice. It can hence not be absorbed into the
  matching as before.
\end{itemize}
The decisive trick to solve both problems can be learned from \
{\cite{Samuel:1978zx}}. We introduce two fields $\xi_{\mu} (x)$ either of
which has two spinor components. Now $\xi_0$ has hopping terms in the zero
direction only and $\xi_1$ implements the perpendicular hops. Our Ansatz for a
bilinear action is\footnote{The unusual sign of $\kappa$ will turn out to be
convenient later.}
\begin{equation}
  S = \sum_x s_0 (\xi_{\mu} (x)) + \kappa \sum_{x, \mu} \bar{\xi}_{\mu}
  (x) P (\hat{\mu}) \xi_{\mu} (x + \hat{\mu}) \label{SMq}
\end{equation}
with unit vectors $\hat{\mu}$ pointing in the positive $\mu$ direction. To
determine the on-site term $s_0$ we postulate
\begin{equation}
  \int d^4 \xi \mathe^{- s_0} \xi_{0, \alpha} \bar{\xi}_{0, \beta} = \int d^4
  \xi \mathe^{- s_0} \xi_{1, \alpha} \bar{\xi}_{1, \beta} = \delta_{\alpha
  \beta}
\end{equation}
to connect straight sections, and
\begin{equation}
  \int d^4 \xi \mathe^{- s_0} \xi_{0, \alpha} \bar{\xi}_{1, \beta} \equiv \int
  d^4 \xi \mathe^{- s_0} \xi_{1, \alpha} \bar{\xi}_{0, \beta} = \sqrt{2}
  \delta_{\alpha \beta}
\end{equation}
to cancel the corner weights. A short calculation shows that this is uniquely
achieved by the quadratic form
\begin{equation}
  s_0 = \frac{1}{2} (\bar{\xi}_0 \xi_0 + \bar{\xi}_1 \xi_1) - \sqrt{2} 
  \bar{\xi}_0 \xi_1 .
\end{equation}
A novelty arises for the empty sites. They now contribute minus signs to the
loop amplitude because we find
\begin{equation}
  \int d^4 \xi \mathe^{- s_0} = - 1. \label{emptyminus}
\end{equation}
The integral with all four Grassmann components is now determined and reads
\begin{equation}
  \int d^4 \xi \mathe^{- s_0} \xi_{0 \alpha}  \bar{\xi}_{0 \beta} \xi_{1
  \gamma}  \bar{\xi}_{1 \delta} = \delta_{\alpha \beta} \delta_{\gamma
  \delta} . \label{4xi}
\end{equation}
We find that at crossings the `vertical' pair gets connected by spin
contraction as well as the `horizontal' one, see figure \ref{cross}. Hence in this
case we now do find separate closed loops, with intersections (including
self-intersections of the same loop) allowed.

The partition function at fixed boundary conditions $\varepsilon$ now reads
\begin{equation}
  Z_M (\kappa, L_{\mu}, \varepsilon) = \int D \xi \mathe^{- S} =
  \sum_{\Lambda} \kappa^{| \Lambda | \nobracket \nobracket} (-
  1)^{\varepsilon_0 q_0 [\Lambda] + \varepsilon_1 q_1 [\Lambda]} (2
  \delta_{q_0 [\Lambda], 0} \delta_{q_1 [\Lambda], 0} - 1) . \label{ZM}
\end{equation}
The sign (\ref{emptyminus}) has been absorbed here into $D \xi = \prod_x (-
d^4 \xi (x))$ to adhere to the normalization $Z_M (0, L_{\mu}, \varepsilon) =
1$. This however implies now extra signs at all non-empty sites, i.e.
connections as well as the crossings (\ref{4xi}). In addition the hopping
terms come with factors $(- \kappa)$. In this way for a graph {\tmem{without}}
crossings, which visits the same number $| \Lambda | \nobracket \nobracket$ of
links and sites, these signs cancel. For each crossing there first seems an
extra minus left over. If a crossing is a self-intersection, this extra sign
cancels however with an extra $2 \pi$ rotation collected along the
corresponding line, which, if it does not run around the torus, then
contributes a total plus sign. If the crossing is between separate loops,
there always is an even number of them.

\begin{figure}[htb]
\begin{center}
  \resizebox{0.5\textwidth}{!}{\includegraphics{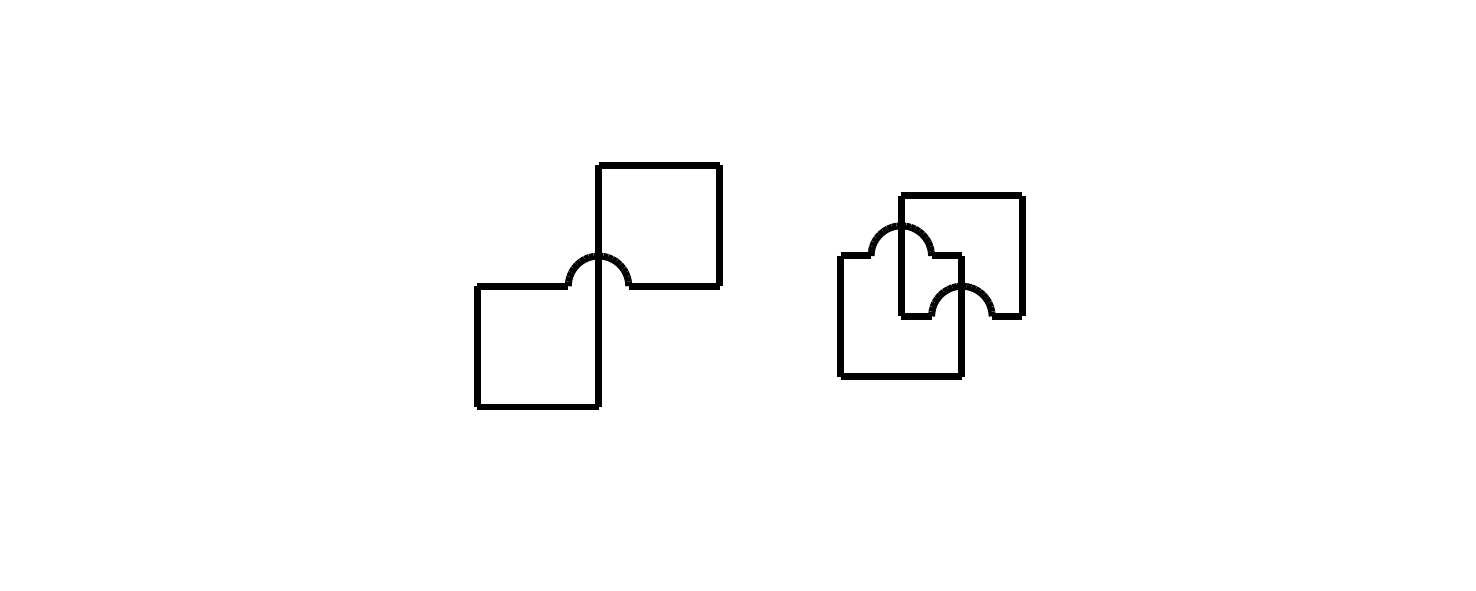}}
  \caption{Examples of self-intersection and intersections of separate
  loops.\label{loops}}
\end{center}
\end{figure}

\

{\noindent}We try to visualize this in figure \ref{loops}. In this way the
number of crossings $n_+ [\Lambda]$ does not appear in the final weight, which
is essential to be able to match the Ising loop gas. The condition for this is
given by
\begin{equation}
  \tanh \beta = \kappa
\end{equation}
for the square lattice. The relation (\ref{Zrel}) between partition functions
holds unchanged. The factor in the last bracket in (\ref{ZM}) has the same
reason as discussed for the honeycomb lattice. We finally re-emphasize that
the loop configurations $\Lambda$ in (\ref{ZM}) are the same as those
described in section (\ref{Isquare}). The exact evaluation of the partition
function (\ref{ZM}) by performing the Gaussian Grassmann integral is discussed
in appendix \ref{appb2}.

If both $L_0$ and $L_1$ are even, also the square lattice is bi-partite and
partition functions are even in $\beta$ or $\kappa$ respectively. If we also
allow for odd lattice lengths the generalized relation
\begin{equation}
  Z_M (- \kappa, L_{\mu}, \varepsilon) = Z_M (\kappa, L_{\mu},
  \varepsilon^{(L)})
\end{equation}
can be proven with
\begin{equation}
  \varepsilon^{(L)}_{\mu} = \varepsilon_{\mu} + L_{\mu} \quad \tmop{mod} 2,
  \label{epsL}
\end{equation}
i. e. a swap between periodic and antiperiodic for odd $L_{\mu}$ directions.

For an easier interpretation of the continuum limit of the action ($\left.
\ref{SMq} \right)$ we diagonalize $s_0$ by changing to new fields $\eta$ and
$\chi$
\begin{equation}
  \xi_0 = \frac{i}{\sqrt{\kappa}}  (\eta + \chi), \qquad \xi_1 =
  \frac{i}{\sqrt{\kappa}}  (\eta - \chi) .
\end{equation}
Introducing the forward, backward and symmetric difference operators
$\partial_{\mu}$, $\partial_{\mu}^{\ast}$ and $\tilde{\partial}_{\mu}$, the
action reads
\begin{eqnarray}
  S & = & \frac{1}{2} \sum_x \bar{\eta} \left( m_{\eta} + \gamma_{\mu}
  \tilde{\partial}_{\mu} - \frac{1}{2} \partial_{\mu}
  \partial_{\mu}^{\ast} \right) \eta + \frac{1}{2} \sum_x \bar{\chi} \left(
  m_{\chi} + \gamma_{\mu} \tilde{\partial}_{\mu} - \frac{1}{2}
  \partial_{\mu} \partial_{\mu}^{\ast} \right) \chi \nonumber\\
  &  & + \sum_x \bar{\eta} \left( \gamma_0 \tilde{\partial}_0 - \gamma_1
  \tilde{\partial}_1 - \frac{1}{2} \partial_0 \partial_0^{\ast} +
  \frac{1}{2} \partial_1 \partial_1^{\ast} \right) \chi 
\end{eqnarray}
with
\begin{equation}
  m_{\eta} = \frac{2}{\kappa} \left[ \sqrt{2} - 1 - \kappa \right], \qquad
  m_{\chi} = - \frac{2}{\kappa} \left[ \sqrt{2} + 1 + \kappa \right] .
\end{equation}
The standard Ising critical point appears at
\begin{equation}
  m_{\eta} = 0 \quad \Leftrightarrow \quad \kappa = \kappa_c = \tanh \beta_c =
  \sqrt{2} - 1 \quad \Rightarrow \quad \beta_c = \frac{1}{2} \ln \left(
  \sqrt{2} + 1 \right)
\end{equation}
While the field $\eta$ is critical here and acquires long range correlations
we have a large negative mass in lattice units $m_{\chi} = - 4 \left( 2 +
\sqrt{2} \right)$. Hence the coupling to this field only contributes small
lattice corrections to the Euclidean symmetric Majorana `particles' described
by $\eta$. As before small positive $m_{\eta}$ ($\kappa < \kappa_c$)
corresponds to the symmetric Ising phase with the ferromagnetic one being on
the other side at negative $m_{\eta}$ ($\kappa > \kappa_c$).

The two fields swap their roles at
\begin{equation}
  m_{\chi} = 0 \quad \Leftrightarrow \quad \kappa = \kappa_c' = \tanh \beta_c'
  = - \sqrt{2} - 1 \quad \Rightarrow \quad \beta_c' = - \beta_c \pm i
  \frac{\pi}{2}
\end{equation}
with $m_{\eta} = - 4 \left( 2 - \sqrt{2} \right)$ in this case.

\subsection{Relation with reference \cite{Samuel:1978zx}}

In \cite{Samuel:1978zx} Samuel has employed Grassmann variables to reproduce
the low temperature expansion of the Ising model (Bloch walls) in powers of\footnote{We take $z_h=z_v=z$ for simplicity.}
$z=\exp(-2\beta)$. As the 2-dimensional model on the square lattice is self-dual, 
this coincides with the high temperature $\tanh \beta$ expansion (finite size effects
are disregarded in \cite{Samuel:1978zx}).

In a first step we adapt Samuels notation for the Grassman fields by replacing
\begin{equation}
  (\eta^{h^x},-\eta^{h^o},\eta^{v^x},-\eta^{v^o}) \to (\eta_{01},\eta_{02},\eta_{11},\eta_{12}),
\end{equation}
and temporarily assume gamma matrices $\gamma_0=\tau_3$, $\gamma_1=\tau_1$ in terms of Pauli matrices. 
The action (3.4) in \cite{Samuel:1978zx} for the Ising case now translates to
\begin{equation}
A=-z \sum_{x\mu}\bar{\eta}_{\mu}(x)P(\hat{0}) \eta_{\mu}(x+\hat{\mu}) -
\frac12 \sum_{x\mu} \bar{\eta}_{\mu} \eta_{\mu} +\sum_x \bar{\eta}_0 (1+\mathcal{C}^{-1}) \eta_1,
\end{equation}
where spin summations are implicit again. To bring the hopping terms into the same form as in (\ref{SMq}),
we perform a spinor rotation $\eta_1 \to R \eta_1$ with $R=\exp(i\pi \tau_2/4)$ which yields
$R^{\dag} \gamma_0 R = \gamma_1$. In terms of these fields the total action now reads
\begin{equation}
A=-z \sum_{x\mu}\bar{\eta}_{\mu}(x)P(\hat{\mu}) \eta_{\mu}(x+\hat{\mu}) -
\frac12 \sum_x \bar{\eta} \eta +\sum_x \bar{\eta}_0 (1+\mathcal{C}^{-1})R \eta_1.
\end{equation}
Using now $\mathcal{C}=i\tau_2$ and $R=(1+i\tau_2)/\sqrt{2}$ we find complete agreement
with the manifestly (cubic) rotation invariant form (\ref{SMq}).

\section{Conclusions and outlook}\label{sec4}

We have given an exact mapping between the Ising model and free Majorana
Wilson fermions for finite honeycomb and square lattices. The critical point
occurs at vanishing mass $m$ or, equivalently, the critical hopping parameter
$\kappa_c$. Although trivial, we mention that the equivalence with free
fermions immediately explains the value $\nu = 1$ for the correlation length
exponent as this scale is given by the inverse mass. The magnetized phase with
broken $Z (2)$ symmetry corresponds to $\kappa > \kappa_c$ or negative mass.

The question of extensions to three dimensions comes to mind. There is a
lattice with coordination number three, the so-called hydrogen-peroxide
lattice {\cite{Liu:2012ca}}. The loop expansion of the three dimensional
Majorana Wilson fermion worked out in {\cite{Wolff:2008xa}} for the cubic
lattice can be adapted to this case by just eliminating a fraction of the links.
Then the graphs `drawn' by the free fermions would indeed coincide with those
of the $\tanh \beta$ expansion of the Ising model, namely a gas of
non-intersecting closed loops. However, as explicitly worked out in
{\cite{Wolff:2008xa}}, any such fermion graph containing non-planar loops
comes with spin phase factors in the group $Z (8)$ -- related to cubic lattice
rotations -- which oscillate in an essential way. Therefore, the graph weights
{\tmem{cannot}} be matched in this case.

\appendix\section{Spin factor\label{appa}}

The calculation in this appendix closely follows the arguments given in
appendix B of {\cite{Wolff:2008xa}}, but is generalized here beyond the square
lattice.

We consider a single closed loop $\lambda$ of length $N$ on a 2 dimensional
lattice to be associated with a sequence of lattice unit vectors $n_i, i = 1,
2, \ldots, N$ which connect nearest neighbors and add to zero
\begin{equation}
  \sum_{i = 1}^N n_i = 0.
\end{equation}
For a given starting point $x_0$ on the lattice, all points recursively given
by\\
$x_i = x_{i - 1} + n_i$ are nearest neighbor lattice sites until the loop
closes at $x_N = x_0$. The spin factor associated with such a loop $\lambda$
is given by the traced product of Wilson projectors (\ref{wdef}) where the
additional Fermi minus is included. Note that $w$ is invariant under cyclic
changes of the $n_i$ and under inversions due to (\ref{Ctrafo}). Hence neither
the starting point along the loop nor the chosen orientation matters for $w$,
which thus is a function of the unoriented loop.

For the evaluation of $w$ we note each pair of $n_i, n_j$ can be rotated into
each other. Using the spinor representation this allows us to write
\begin{equation}
  P (n_{i + 1}) = R_i^{- 1} P (n_i) R_i
\end{equation}
with
\begin{equation}
  R_i = \exp \left( \frac{\alpha_i}{2} \gamma_0 \gamma_1 \right) \quad
  \tmop{with} \quad \cos (\alpha_i) = n_i \cdot n_{i + 1} .
\end{equation}
This is used, starting from the rightmost factors in the product,
\begin{equation}
  P (n_{N - 1}) P (n_N) = P (n_{N - 1}) R_{N - 1}^{- 1} P (n_{N - 1}) R_{N -
  1} = \cos (\alpha_{N - 1} / 2) P (n_{N - 1}) R_{N - 1}^{}
\end{equation}
where we have used $P \exp (\alpha \gamma_0 \gamma_1) P = \cos (\alpha) P$.
Upon iteration we arrive at
\begin{equation}
  w (\lambda) = - \left\{ \prod_{i = 1}^{N - 1} \cos (\alpha_i / 2)  \right\}
  \tmop{tr} [P (n_1) R_1 R_2 \cdots R_{N - 1}] .
\end{equation}
If we define the additional rotation $R_N$ to achieve
\begin{equation}
  P (n_1) = R_N^{- 1} P (n_N) R_N,
\end{equation}
then the total rotation
\begin{equation}
  R_{\lambda} = R_1 R_2 \cdots R_{N - 1} R_N
\end{equation}
has the direction $n_N$ as a fixed point
\begin{equation}
  P (n_N) = R_{\lambda}^{- 1} P (n_N) R_{\lambda}
\end{equation}
which implies
\begin{equation}
  R_{\lambda} = \exp \left( \gamma_0 \gamma_1 \frac{1}{2} \sum_{i = 1}^N
  \alpha_i \right) = \pm 1
\end{equation}
and
\begin{equation}
  w (\lambda) = - R_{\lambda} \prod_{i = 1}^N \cos (\alpha_i / 2) .
  \label{weval}
\end{equation}
For simple non-selfintersecting contractable closed loops in the plane the
angles $\alpha_i$ add up to $2 \pi$ and thus $R_{\lambda} = - 1$ holds. This
clearly is the minus sign under a $2 \pi$ rotation that a spinor receives as
it is transported once around the closed loop. Note that this does not occur
for loops closing around the torus in one or both directions. Bends along the
loops are suppressed by weight factors $\cos (\pi / 6) = \sqrt{3} / 2$
(honeycomb) and \ $\cos (\pi / 4) = 1 / \sqrt{2}$ (square).

\section{Exact dispersion of Wilson fermions}\label{appb}

\subsection{Honeycomb lattice}\label{appb1}

We switch to sublattice Majorana fields
\begin{equation}
  \chi (x) = (\chi_A (x), \chi_B (x)) \equiv (\xi (x), \xi (x + \hat{e}_0))
\end{equation}
with the 4-component field $\chi$ attached to sublattice $\mathcal{A}$. We write
down a Fourier representation
\begin{equation}
  \chi (x) = \frac{1}{L_1 L_2}  \sum_p \tilde{\chi} (p) \mathe^{i (p_1 x_1 +
  p_2 x_2)} . \label{Four}
\end{equation}
Some straightforward algebra yields the action (\ref{Smajo}) in terms of
$\tilde{\chi}$
\begin{equation}
  S = \frac{1}{2 L_1 L_2}  \sum_p \{ \bar{\tilde{\chi}} (- p)
  \tilde{\chi} (p) - \kappa [\bar{\tilde{\chi}}_A (- p) M_+
  \tilde{\chi}_B (p) + \bar{\tilde{\chi}}_B (- p) M_- \widetilde{\chi_A}
  (p)] \} \label{Smajop}
\end{equation}
with
\begin{equation}
  M_{\pm} (p) = \sum_{j = 0}^2 P (\pm \hat{e}_j) \mathe^{\pm i p_j} \qquad
  (p_0 \equiv 0) .
\end{equation}
The momenta to be summed over depend on the lattice size $L_i$ and
(anti)periodicity $\varepsilon_i$. A possible choice would be $p_i = (2 \pi /
L_i) (\varepsilon_i / 2 + n_i), n_i = 0, \ldots, L_i - 1$. The special values
$p_i = 0$ are allowed for $\varepsilon_i = 0$ and $p_i = \pi$ occurs if $L_i +
\varepsilon_i$ is even. If all components are of this type, $p$ and $- p$ are
identical\footnote{$p_i$ differing by multiples of $2 \pi$ are identified, of
course.} and so are $\tilde{\chi} (p)$ and $\tilde{\chi} (- p)$. The Grassmann
integral for such momenta then leads to a Pfaffian of the quadratic form
defined by (\ref{Smajop}). The remaining momenta come in pairs associated with
independent Grassmann fields and contribute determinant factors to the
partition function, one per pair. We divide up \ the set of all momenta as
follows,
\begin{equation}
  \mathcal{B} (L_i, \varepsilon_i) =\mathcal{B}_0 (L_i, \varepsilon_i) \cup
  \mathcal{B}_+ (L_i, \varepsilon_i) \cup \mathcal{B}_- (L_i, \varepsilon_i),
\end{equation}
where $\mathcal{B}_0$ contains momenta made of components $0$ or $\pi$ only,
while $\mathcal{B}_+$ contains {\tmem{one member}} of each of the remaining
pairs $\pm p$ with the partner momenta in $\mathcal{B}_-$. This implies for
the cardinalities $| \mathcal{B}_0 | + 2 | \mathcal{B}_+ | = L_1 L_2
\nobracket \nobracket \nobracket \nobracket$ to hold.

In any case we may perform half of the Gaussian integrations - say over
$\tilde{\chi}_B$ - \ to obtain the reduced action
\begin{equation}
  S_A = \frac{1}{2 L_1 L_2}  \sum_p \left\{ \bar{\tilde{\chi}}_A
  (- p) \{ 1 - \kappa^2 M_+ M_- \} \tilde{\chi}_A (p) \right\}
\end{equation}
which leads to $2 \times 2$ determinants and Pfaffians. We expand
\begin{equation}
  M_+ M_- = a + i b_{\mu} \gamma_{\mu} + i c \gamma_0 \gamma_1
\end{equation}
and find in a short calculation
\begin{eqnarray}
  a & = & \frac{3}{4} \{ \cos (p_1) + \cos (p_2) + \cos (p_1 - p_2) \}, \\
  b & = & \frac{1}{2} \{ \sin (p_1) f_1 + \sin (p_2) f_2 + \sin (p_1 - p_2)
  (f_1 - f_2) \}, \\
  c & = & \frac{\sqrt{3}}{4} \{ - \sin (p_1) + \sin (p_2) + \sin (p_1 - p_2)
  \} . 
\end{eqnarray}
This implies for the determinants
\begin{equation}
  D (p) = (1 - \kappa^2 a)^2 + \kappa^4 (b_{\mu} b_{\mu} - c^2) .
\end{equation}
For momenta in $\mathcal{B}_0$ the contributions $b_{\mu}$ and $c$ vanish and
the Pfaffian is given by
\begin{equation}
  P (p) = 1 - \kappa^2 a \qquad (p_{\mu} \in \{ 0, \pi \}) .
\end{equation}
Note that $P^2 = D$ holds here, but the root of $D$ has to be taken such, that
$P$ is a polynomial in $\kappa$. In total we arrive at
\begin{equation}
  Z_M (\kappa, L_i, \varepsilon_i) = \left\{ \prod_{p \in \mathcal{B}_0 (L_i,
  \varepsilon_i)} P (p) \right\} \prod_{p \in \mathcal{B}_+ (L_i,
  \varepsilon_i)} D (p) .
\end{equation}

\begin{figure}[htb]
\begin{center}
  \resizebox{0.8\textwidth}{!}{\includegraphics{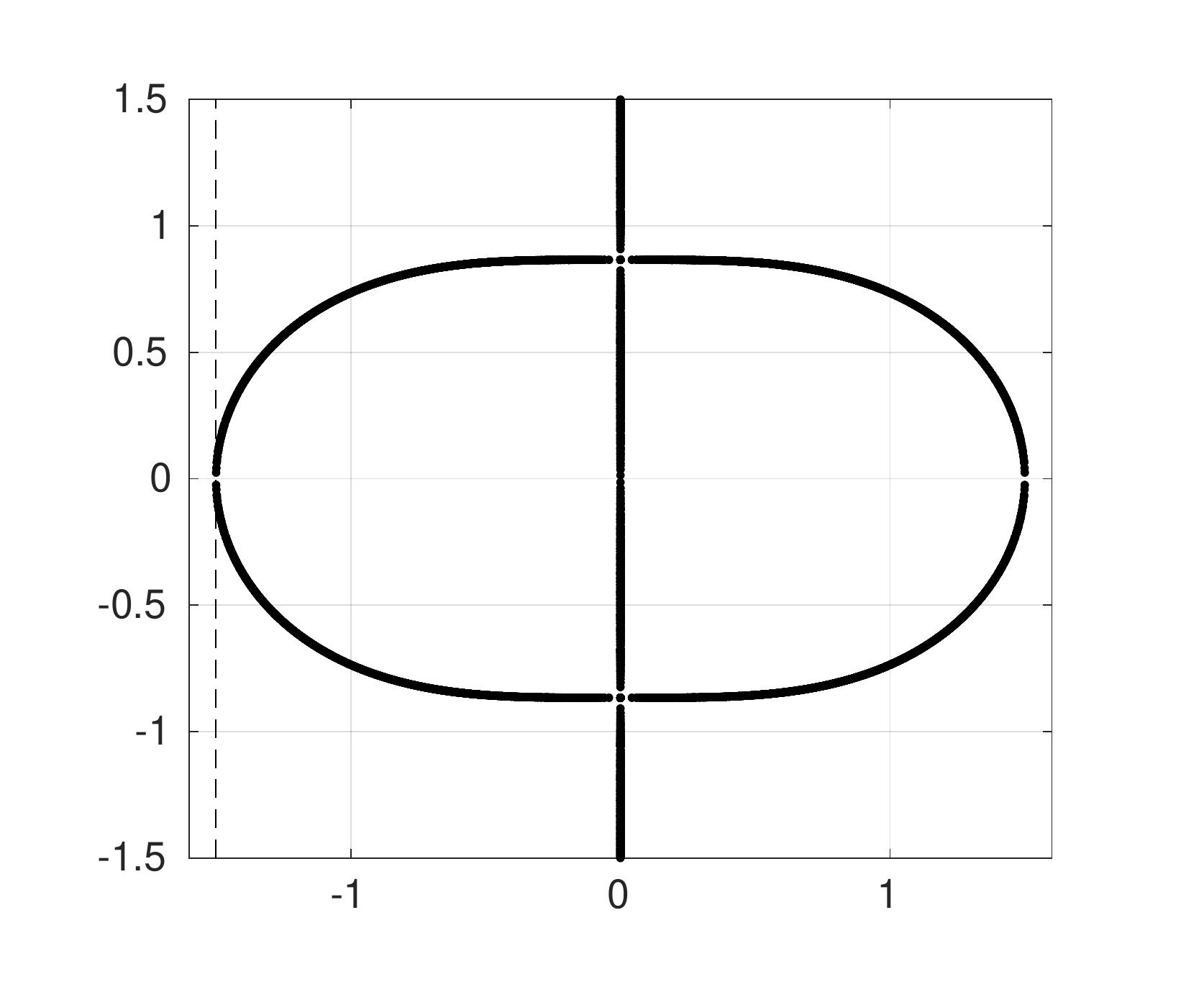}}
  \caption{Fermion spectrum $\rho$ for $L_1 = L_2 = 64$ and antiperiodic
  boundary conditions.\label{specma}}
\end{center}
\end{figure}

The four eigenvalues $\lambda = 1 + \kappa \rho$ (for each $p$) of the
quadratic form (\ref{Smajop}), which control the fermion two-point function, \
are given by
\begin{equation}
  \rho = \pm \sqrt{a \pm \sqrt{c^2 - b_{\mu} b_{\mu}}} \quad (4 \tmop{sign}
  \tmop{combinations}) .
\end{equation}
The spectrum of complex $\rho$ values is shown in figure \ref{specma}. The
spectral radius $3 / 2$ corresponds to the critical values $\kappa_c = 2 / 3$.
The spectrum is invariant under $\rho \rightarrow \rho^{\ast}$ as one can show
$M_{\pm}^{\ast} =\mathcal{C}M_{\mp} \mathcal{C}^{- 1}$, and under $\rho
\rightarrow - \rho$ that is related to the\\
$\kappa \rightarrow - \kappa$ symmetry. The arc of eigenvalues tangent to the
dashed line in figure \ref{specma} approximates the imaginary spectrum of the
continuum Euclidean Dirac operator $\gamma_{\mu} \partial_{\mu}$.

Upon expanding for small $p_1, p_2$ one finds that $a$ and $b_{\mu} b_{\mu}$
depend on the combination $p_1^2 + p_2^2 - p_1 p_2$ while $c$ only contributes
to higher orders. To arrive at the Fourier form (\ref{Four}) we actually
expand $p = p_1 \widetilde{f_1} + p_2 \widetilde{f_2}$ in the basis dual to
(\ref{fsp}) which is defined by $f_i \cdot \tilde{f}_j = \delta_{i j}$. Then
by elementary steps we find
\begin{equation}
  p^2 = \frac{4}{9} (p_1^2 + p_2^2 - p_1 p_2) .
\end{equation}
Hence this combination is Euclidean invariant in our basis and so is the free
energy and the dispersion in the continuum limit.
\subsection{Square lattice}\label{appb2}

\begin{figure}[htb]
\begin{center}
  \resizebox{0.5\textwidth}{!}{\includegraphics{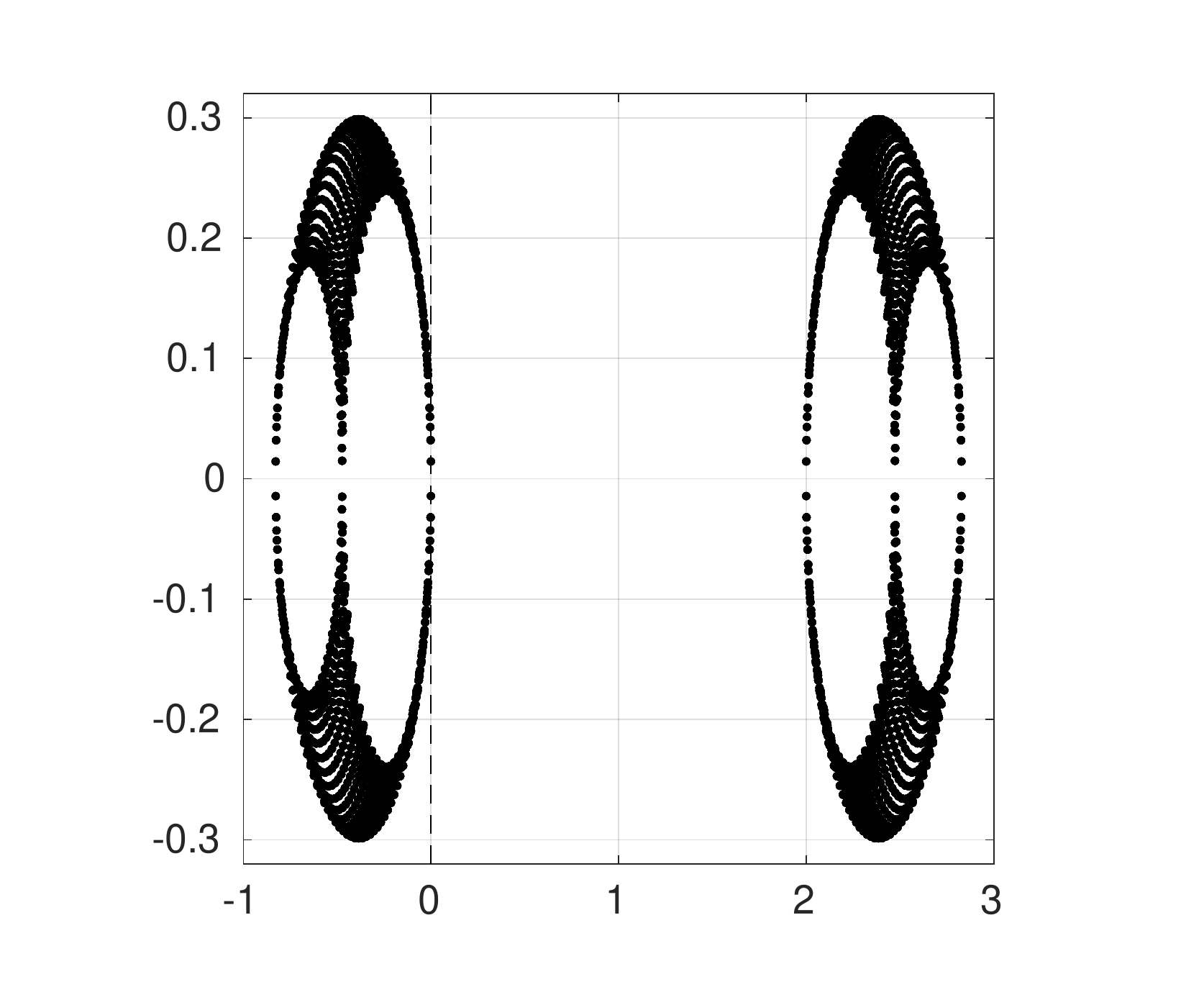}}\resizebox{0.5\textwidth}{!}{\includegraphics{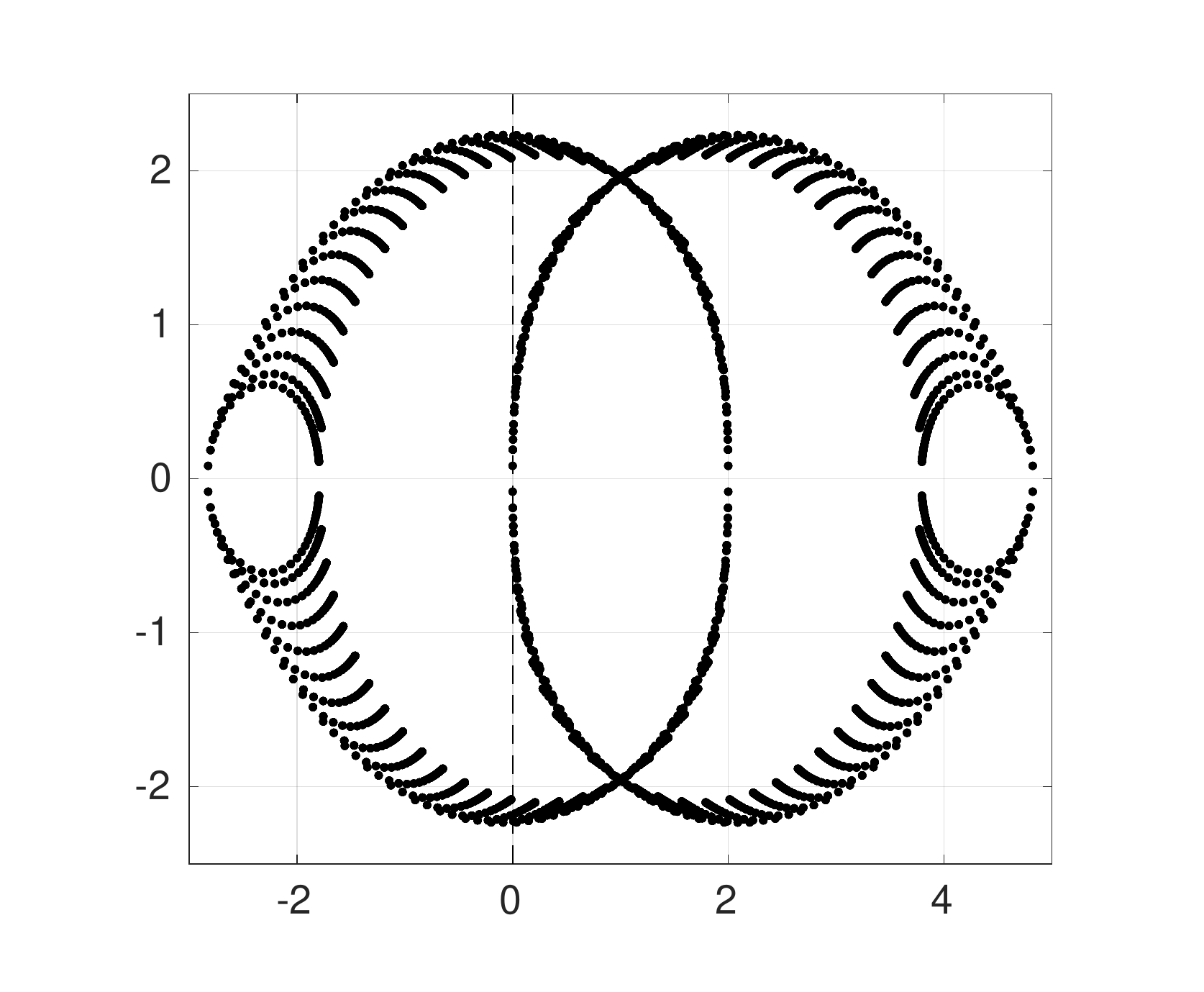}}
  \caption{Eigenvalues $\lambda$ for $L_{\mu} = 64$ and antiperiodic boundary
  conditions
  for $\kappa = \sqrt{2} - 1$ (left plot) and $\kappa = - \sqrt{2} - 1$
  (right plot).\label{specmaq}}
\end{center}
\end{figure}

The action in momentum space can be written as
\begin{equation}
    S  =  \frac{1}{2 L_0 L_1}  \sum_p \left\{ \sum_{\mu}
    \bar{\tilde{\xi}}_{\mu} (- p) [1 + \kappa \mathe^{- i p_{\mu}
    \gamma_{\mu}}] \tilde{\xi}_{\mu} (p) - \sqrt{2} [\bar{\tilde{\xi}}_0
    (- p) \tilde{\xi}_1 (p) + \bar{\tilde{\xi}}_1 (- p) \tilde{\xi}_0
    (p)] \right\}
    \label{Smasq}
\end{equation}
By similar manipulations as in the previous subsection we may work out the
characteristic polynomial whose zeros are the eigenvalues of the quadratic
form defined by (\ref{Smasq}),
\begin{equation}
  C (\lambda) = \sum_{i = 0}^4 c_i (1 - \lambda)^i,
\end{equation}
with
\begin{eqnarray*}
  c_4 & = & 1\\
  c_3 & = & 2 \kappa [\cos (p_0) + \cos (p_1)]\\
  c_2 & = & 2 \kappa^2 [1 + 2 \cos (p_0) \cos (p_1)] - 4\\
  c_1 & = & 2 \kappa (\kappa^2 - 2) [\cos (p_0) + \cos (p_1)]\\
  c_0 & = & \kappa^4 + 4 [1 - \kappa^2 \cos (p_0) \cos (p_1)]
\end{eqnarray*}
While closed expressions for the eigenvalues can be given now, we found them
not very illuminating and content ourselves with figure \ref{specmaq} for
$\kappa = \pm \sqrt{2} - 1$. The dashed vertical lines are tangent to
approximate continuum spectra of $\gamma_{\mu} \partial_{\mu}$. One shows
invariance of the spectrum under $\lambda \rightarrow \lambda^{\ast}$ due to the
property ($\nobracket \mathe^{- i p_{\mu} \gamma_{\mu}})^{\ast} =\mathcal{C}
\mathe^{- i p_{\mu} \gamma_{\mu}} \mathcal{C}^{- 1}$. The symmetry $\lambda
\rightarrow 2 - \lambda$ holds for even $L_{\mu}$ or requires a simultaneous
change in the boundary conditions as in (\ref{epsL}).

To compute $Z_M$ the momenta are divided as in the previous subsection and the
Grassmann integrations again lead to Pfaffians and determinants. The result is
\begin{equation}
  Z_M (\kappa, L_{\mu}, \varepsilon_{\mu}) = \left\{ \prod_{p \in
  \mathcal{B}_0 (L_{\mu}, \varepsilon_{\mu})} P (p) \right\} \prod_{p \in
  \mathcal{B}_+ (L_{\mu}, \varepsilon_{\mu})} D (p)
\end{equation}
with
\begin{equation}
  D (p) = C (0) = (1 + \kappa^2)^2 + 2 \kappa (\kappa^2 - 1) [\cos (p_0) +
  \cos (p_1)]
\end{equation}
and for $p \in \mathcal{B}_0$ we find the Pfaffians
\begin{equation}
  P (p) = \left\{\begin{array}{lll}
    2 - (1 + \kappa)^2 & \tmop{for} & p = (0, 0)\\
    1 + \kappa^2 & \tmop{for} & p = (\pi, 0), (0, \pi)\\
    2 - (1 - \kappa)^2 & \tmop{for} & p = (\pi, \pi)
  \end{array}\right. .
\end{equation}
For small $p_{\mu}$ we obviously find the dependence on the relativistic
invariant $p_0^2 + p_1^2$ at leading order in all terms.

\end{document}